\documentclass[final,5p,times,twocolumn]{elsarticle}

\usepackage{graphicx}%
\usepackage{amsmath,amssymb,amsfonts}%
\usepackage{xcolor}%
\usepackage{enumitem}%
\usepackage{physics}%
\usepackage{comment}%

\usepackage{hyperref}
\hypersetup{
	colorlinks=true,
	linkcolor=blue,
	filecolor=magenta,      
	urlcolor=cyan,
	pdftitle={Overleaf Example},
	pdfpagemode=FullScreen,
}

\begin{document}

	\begin{frontmatter}

		\title{A Non-Hermitian Potential Well Formalism for \\ Conscious--Preconscious--Subliminal Processing}

		\author{Vasily Lubashevskiy\corref{cor1}} 
		\ead{vlubashe@tiu.ac.jp}
		\cortext[cor1]{Corresponding author}
		\affiliation{organization={Institute for International Strategy, Tokyo International University},
			addressline={4 Chome-42-31 Higashiikebukuro, Toshima}, 
			city={Tokyo},
			postcode={170-0013}, 
			country={Japan}}
	
		\author{Ihor Lubashevsky} 
		\ead{ilubashevskii@hse.ru}
		\affiliation{organization={Tikhonov Moscow Institute of Electronics and Mathematics, HSE University},
			addressline={34 Tallinskaya str.}, 
			city={Moscow},
			postcode={123458}, 
			country={Russia}}

		\begin{abstract}	
		
We propose a phenomenological model of the Global Neuronal Workspace (GNW) in which early sensory processing generates an effective complex-valued landscape governing the dynamics of high-level stimulus representations. This landscape provides a dynamical bridge between sensory encoding and conscious access, enabling both processes to be described within a unified framework. High-level representations are encoded in a cloud function defined on a Hilbert space over a perceptual state space, thereby combining the holistic structure of mental images with a neural implementation. Its dynamics is governed by a nonlinear Schr\"odinger-type equation in imaginary time with a non-Hermitian, non-normal Hamiltonian and a nonlinear Lotka--Volterra-type term that preserves norm and enables spatially nonlocal interactions. The Hermitian and anti-Hermitian parts of the Hamiltonian generate complementary processes: recognition via dissipative localization at minima of the GNW landscape and information broadcasting via spatial spreading across the state space. The resulting dynamics reproduces the subliminal--preconscious--conscious hierarchy of sensory processing. Conscious access corresponds to the emergence of a bound state, which occurs only when both the GNW landscape depth and the degree of top-down attention exceed threshold values. The resulting framework provides a tractable dynamical description linking sensory processing, attention, and conscious access within a unified dynamical setting.
		
		\end{abstract}

		\begin{keyword}
			Neural field theory \sep
			Global neuronal workspace \sep
			Complex-valued landscape \sep
			Non-Hermitian Schr\"odinger equation \sep
			Bound states \sep
			Sensory processing  
			
			
			
			
		\end{keyword}
		
	\end{frontmatter}

\section{Introduction}\label{sec1}
 
The Global Neuronal Workspace (GNW) hypothesis, originally proposed by Dehaene \textsl{et al.}~\citep{Dehaene1998}, has become one of the leading theories of consciousness \citep[][for reviews]{Mashour2020,Ferrante2025}; see also \citep{Seth_2022,Mudrik2025} for its place among contemporary theories of consciousness. According to the GNW theory, a stimulus reaches consciousness only when two conditions are satisfied. First, the neural activity elicited during the early stage of sensory processing must be sufficiently strong to exceed the threshold for global ignition, thereby triggering large-scale recurrent (reverberating) neural activity. Second, this activity must be amplified by top-down attention. Consequently, conscious access is prevented when attention is directed toward another stimulus or task. These principles give rise to the following taxonomy of information processing in the brain~\citep{Dehaene2006,Kouider_2007}:
\begin{itemize}[topsep=0.1\baselineskip,itemsep=0.1\baselineskip,label=--]
	\item \textit{Subliminal processing}, in which neural activity is too weak for bottom-up activation alone to trigger global ignition. Consequently, the corresponding sensory information cannot reach consciousness, regardless of the allocation of top-down attention.
	\item \textit{Preconscious (supraliminal unattended) processing}, in which conscious access is limited by the availability of top-down attention rather than by the strength of bottom-up activation. Although the neural activity is sufficiently strong to support global ignition, the corresponding information is temporarily maintained in a nonconscious buffer because top-down attention is unavailable.
	\item \textit{Conscious (supraliminal attended) processing}, in which global ignition is achieved and the corresponding neural representation is broadcast throughout the GNW, making the associated information available to multiple specialized processing systems.
\end{itemize}
Moreover, early sensory processing and conscious processing constitute distinct dynamical regimes: the former is dominated by transient feedforward activity, whereas the latter is characterized by nonlinear ignition and sustained recurrent interactions within the GNW~\citep{Mashour2020}. This distinction naturally motivates viewing the GNW as a multilevel model of conscious processing~\citep{Changeux2026}.
 
We recently proposed a phenomenological description of sensory processing \citep{LubashNatPhen2025}, in which the early and late (high-level) stages are treated as distinct levels, each characterized by its own stimulus representations. Their interaction is mediated by an intermediate bridge that gives rise to an effective landscape. This landscape is shaped by early sensory processing and, in turn, governs the dynamics of high-level representations. On the one hand, the proposed framework captures the central principles of the predictive coding paradigm \citep[e.g.,][]{friston.rstb.2005,clark2013whatever}. On the other hand, both the high-level representations and the landscape are assumed to be reconstructed from the first-person perspective, that is, from the mental images of perceived objects. In particular, the spatial organization of high-level representations is assumed to inherit that of the corresponding mental images---space--time clouds \citep{LubPlav2021}. This observation motivated us to introduce the concept of cloud functions, whose dynamics is identified with sensory processing. The resulting formalism enabled us, in particular, to propose a novel explanation for the power law of working memory (PLWM) \citep[][for a review of PLWM]{Smith2018}.

We subsequently generalized this framework \citep{Lubashevskiy2026} by interpreting cloud functions as a special class of complex-valued neural fields and the GNW as a Hilbert space whose elements are cloud functions. The effective landscape governing the evolution of these functions, referred to as the GNW landscape, serves as the dynamical interface linking early sensory processing, where it is formed, to the recurrent dynamics of cloud functions within the GNW. This formalism was then used to propose a novel explanation for the change-of-mind effect in decision-making, whereby an initially selected option is replaced by an alternative during the execution of the decision \cite{Resulaj2009}.

Concerning top-down attention, we adopt the limited-resource framework of attention, a family of models describing various forms of attentional allocation \citep[][for reviews]{Wickens2021,Denison2024}, together with the normalization model of attention \citep{Reynolds2009} and its extension \citep{Schwedhelm2016}. Specifically, we assume that the allocation of computational resources modulates sensory processing through multiplicative gain of stimulus-driven activity followed by divisive normalization by the surrounding neuronal population. Accordingly, the effective sensory processing rate $R$ is enhanced according to
\[
R \to G(A)R,
\]
where $A$ denotes the attention degree, i.e., the fraction of the available computational resources allocated to processing the stimulus, and $G(A)$ is the corresponding attentional gain. Consistent with normalization models, the attentional gain is generally expected to be a saturating (sigmoidal) function of $A$. Since only the attentional gain enters the formulation developed below, we adopt the linear approximation
\[
G(A)=A,\qquad 0\le A\le 1,
\]
which corresponds to the leading-order behavior of a general saturating dependence. Thus, for convenience, we identify the attention degree with the normalized attentional gain and denote both by $A$. Under this convention, $A=1$ corresponds to the allocation of the maximum available computational resources.

The purpose of the present paper is to demonstrate that the proposed neural field theory naturally reproduces the aforementioned taxonomy of sensory processing. Specifically, minima of the GNW landscape, identified with the stimulus representations established during the early stage of sensory processing, may or may not give rise to cloud-function bound states, depending on the depth of the corresponding minimum and the degree of top-down attention.
 

\section{Model Description}\label{MD}

\subsection{General structure}

As noted in the Introduction, the early stage of sensory processing is assumed to give rise to an effective potential landscape $\Omega(x)$ within the global neuronal workspace (GNW). The GNW is modeled as the Hilbert space $\mathbb{H}=L^2(\mathbb{R}^N)$ over an $N$-dimensional state space $\mathbb{R}^N$, whose points represent possible properties that an observed physical system may possess. The elements of $\mathbb{H}$ are cloud functions $\Psi(x,t)$, which describe the high-level representation of the system and its environment in the GNW at time $t$. The Hilbert space $\mathbb{H}$ is equipped with the standard inner product
\begin{equation}\label{IP:1}
	\bra{\Psi_1}\ket{\Psi_2}
	=
	\int_{\mathbb{R}^N}\Psi_1^*(x)\Psi_2(x)\,dx,
\end{equation}
subject to the normalization condition
\begin{equation}\label{IP:2}
	\int_{\mathbb{R}^N} |\Psi(x,t)|^2\,dx = 1.
\end{equation}
It is important to emphasize that the cloud function $\Psi(x,t)$ encompasses features that are both accessible and inaccessible to consciousness. Furthermore, its nonlocality in $\mathbb{R}^N$ represents perceptual uncertainty arising from the neural mechanisms underlying sensory processing in the brain. In particular, the quantity $|\Psi(x,t)|^2$ is interpreted as a normalized density over perceptual configurations.

The GNW landscape $\Omega(x)$, together with top-down selective attention, governs the evolution of high-level representations within the GNW. Following \citep{LubashNatPhen2025,Lubashevskiy2026}, we describe the dynamics of the cloud function $\Psi$ by the equation
\begin{equation}\label{eq:1}
	\tau\frac{\partial\Psi}{\partial t}
	=
	-\hat{\mathcal{H}}\Psi
	+
	\ev{\hat{\mathcal{H}}}{\Psi}\Psi,
\end{equation}
where $\tau \sim 200$~ms is the characteristic time scale of high-level visual processing within the GNW \citep[e.g.,][]{Mashour2020}, and the priority Hamiltonian $\hat{\mathcal{H}}$ incorporates both the landscape $\Omega(x)$ and the effects of selective attention. It is worth noting that each eigenfunction of the non-Hermitian Hamiltonian $\hat{\mathcal{H}}$ individually represents a steady state, most of which are unstable. Only a small subset of these eigenfunctions are stable or metastable, and their dynamics are associated with the perception of system properties.

As discussed in \citep{Lubashevskiy2026}, the action of the operator $\hat{\mathcal{H}}$ on $\Psi$ can generally be represented as a convolution. Consequently, Eq.~\eqref{eq:1} accounts for interactions between neuronal populations mediated by effective connectivity across the cortical sheet at both short and long spatial scales. Although long-range connectivity plays an essential role in integrative brain function, neuronal interactions are typically stronger locally than over large spatial distances \citep[e.g.,][]{Betzel2018,Sato2021,Wang2022}. Therefore, in the present analysis, we adopt a short-range approximation of the priority Hamiltonian $\hat{\mathcal{H}}$, retaining only two leading terms. The first term, proportional to $\ell^2\nabla_x^2$, accounts for the finite spatial resolution of neural processing. It suppresses sensitivity to variations on scales much smaller than the characteristic length $\ell$, which may be interpreted as the irreducible uncertainty in the perception of system properties. Although $\ell$ is generally tensor-valued, it is treated here as a scalar for simplicity. The second term describes the influence of the GNW landscape and is approximated by the product of $\Omega(x)$ and the cloud function $\Psi$.

To construct a specific form of the priority Hamiltonian $\hat{\mathcal{H}}$, which is a non-normal operator, it is useful to decompose it into its Hermitian and anti-Hermitian components,
\begin{equation}\label{HO:1}
	\hat{\mathcal{H}} = \hat{\mathcal{H}}' + i \hat{\mathcal{H}}'',
\end{equation}
which play distinct roles in the cloud-function dynamics.

\begin{figure}
	\begin{center}
		\includegraphics[width=0.95\columnwidth]{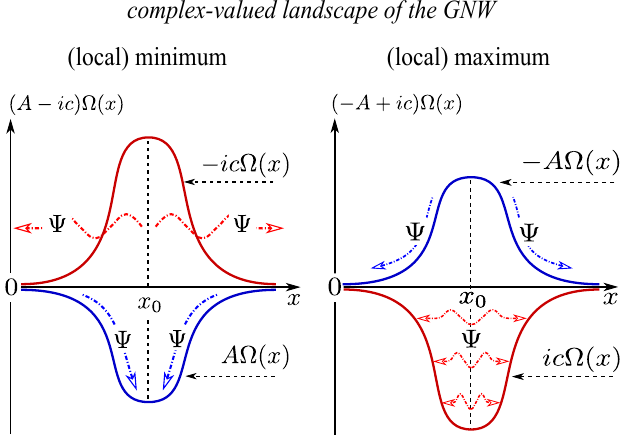}
	\end{center}
	\caption{Schematic illustration of the dual role of the GNW landscape in the dynamics of the cloud function~$\Psi$. A minimum of the GNW landscape is represented by the potential well~$A\Omega(x)$ (blue), whose effective depth is proportional to the attention degree~$A$, and is associated with the recognition process, which drives the cloud function~$\Psi$ toward the minima of~$\Omega(x)$. Information broadcasting across the GNW is associated with the potential barrier~$-ic\Omega(x)$ (red), which counteracts the localization of~$\Psi$. A maximum of the GNW landscape is modeled by reversing the sign of the landscape, $\Omega(x)\to-\Omega(x)$, whereby the potential~$-A\Omega(x)$ promotes the delocalization of~$\Psi$, even though~$ic\Omega(x)$ formally becomes a potential well.}
	\label{F:1}
\end{figure}

We associate the Hermitian component $\hat{\mathcal{H}}'$ with the recognition process driven by top-down attention. This process is interpreted as the localization of the cloud function $\Psi$ near the minima of $\Omega(x)$. Equivalently, $\hat{\mathcal{H}}'$ is assumed to drive $\Psi$ toward these minima, suggesting that the GNW landscape acts as an attractor in the cloud-function dynamics (Fig.~\ref{F:1}). As discussed in the Introduction, selective attention is interpreted as the allocation of limited computational resources within the GNW to process specific sensory information. Accordingly, the contribution of $\hat{\mathcal{H}}'$ to the time evolution of $\Psi$ is taken to be proportional to the attention degree $A$ assigned to a given perceptual item. In other words, we assume $\hat{\mathcal{H}}' \propto A(x,t)$, where $A(x,t)$ represents the relative amount of computational resources allocated to a fragment of the GNW landscape at time $t$.

Combining these considerations, we adopt the ansatz
\begin{equation}\label{HR:1}
	\hat{\mathcal{H}}' = A(x,t)\left[-\ell^2 \nabla_x^2 + \Omega(x)\right].
\end{equation}
In what follows, we consider the case where $A(x,t)$ is taken to be a prescribed constant $A \leq 1$. In other words, we assume that attention is distributed over a region of the GNW landscape that is much broader than the analyzed fragment, as is typically the case during search. The opposite regime requires a separate analysis.

Whereas the Hermitian component $\hat{\mathcal{H}}'$ promotes localization of the cloud function $\Psi$ near the minima of $\Omega(x)$, the anti-Hermitian component $i\hat{\mathcal{H}}''$ promotes its delocalization and is associated with the broadcasting of neural activity across the GNW. In contrast to recognition, this process proceeds without selective attention and remains inaccessible to consciousness. We approximate the operator $i\hat{\mathcal{H}}''$ by the ansatz
\begin{equation}\label{HR:2}
	\hat{\mathcal{H}}'' = \left[-c_\eta\ell^2 \nabla_x^2 - c_\omega\Omega(x)\right],
\end{equation}
where $c_\eta$ and $c_\omega$ are real-valued constants of the same sign, i.e., $c_\eta c_\omega > 0$. With this choice, the minima of the GNW landscape act as effective potential barriers, driving the cloud function $\Psi$ away from these minima (Fig.~\ref{F:1}).

It should be noted that the effect of the GNW landscape maxima can be modeled by reversing the sign of the landscape profile, $\Omega(x)\rightarrow -\Omega(x)$ (Fig.~\ref{F:1}). In this case, the former minimum $A\Omega(x)$ becomes a potential barrier that induces the delocalization of the cloud function $\Psi$, whereas the term $ic_\omega\Omega(x)$ becomes a potential well supporting bound states. As shown below, however, all such states are unstable and, thereby, the maxima of the GNW landscape cannot be associated with the recognition of an external stimulus.

\subsection{$\Psi$-dynamics near a potential well}

Below, we consider a particular case in which the possible properties of an observed system form a one-dimensional state space, $\mathbb{R}$, and the early stage of sensory information processing gives rise to a GNW landscape containing a single potential well. The latter is approximated by the modified P\"oschl--Teller potential \citep[e.g.,][Problem 39]{Cevik2016,Flugge1971},
\begin{equation}\label{eq:Om}
	\Omega(x) = - \frac{U_d}{\cosh^2\big[(x-x_0)/d\big]},
\end{equation}
where $U_d$ is the real-valued well depth, and the parameter $d \ge \ell$ also allows for a finite extent of the perceived property centered at $x_0$. The limiting case $d=\ell$ corresponds to an observed system whose underlying property is point-like and located at $x_0$.

To elucidate the basic features of the proposed mathematical description of the SPC hierarchy, we rewrite the governing equation~\eqref{eq:1} in dimensionless form. In particular, (\textit{i}) time $t$ is measured in units of $(d/\ell)^2\tau$, (\textit{ii}) the new state variable $\eta = (x-x_0)/d$ is introduced, (\textit{iii}) the well depth is rescaled as $U = (d/\ell)^2U_d$, and (\textit{iv}) the parameters  $c_\eta$, $c_\omega$ are set equal, $c_\eta=c_\omega=c$. Such values of $c_\eta$ and $c_\omega$ accentuate the distinction between the roles of the Hermitian and anti-Hermitian components. In these variable, Eq.~\eqref{eq:1} takes the form
\begin{equation}\label{eq:1DL}
	\frac{\partial\Psi}{\partial t}
	=
	\hat{\mathcal{F}}\Psi
	-
	\ev{\hat{\mathcal{F}}}{\Psi}\Psi\,,
\end{equation}
where the operator $\hat{\mathcal F}$ is defined as 
\begin{equation}\label{eq:F1}
	\hat{\mathcal{F}}
	=
	(A+ ic)\frac{\partial ^2}{\partial \eta^2} + (A - ic)\frac{U}{\cosh^2\eta}.
\end{equation}
Provided it exists, the ground-state eigenfunction $\Psi_0$ of the operator $\hat{\mathcal{F}}$ and the corresponding eigenvalue $E_0$---equivalently, the eigenfunction associated with the lowest eigenvalue of the operator $-\hat{\mathcal{F}}$ in its discrete spectrum---are given by
\begin{equation}\label{eq:F2}
	\Psi_0(\eta)
	=
	\frac{Z_0}{\cosh^{\mu}\eta},
	\qquad
	E_0
	=
	(A+ic)\mu^2.
\end{equation}
Here, $Z_0$ is the complex normalization constant whose modulus is fixed by the normalization condition, whereas its phase remains arbitrary. The exponent $\mu$ and the potential depth $U$ are related by
\begin{equation}\label{eq:F3}
	\mu(\mu+1)
	=
	\frac{A-ic}{A+ic}\,U,
	\quad\text{or}\quad
	\mu
	=
	-\frac12
	+
	\sqrt{
		\frac14
		+
		\frac{A-ic}{A+ic}\,U
	}.
\end{equation}
The square root in Eq.~\eqref{eq:F3} is taken on the principal branch, with a branch cut along $(-\infty,0]$.

There are at least two criteria for the stability of the ground-state eigenfunction $\Psi_0$. The first criterion is the existence of $\Psi_0$ itself, which reduces to the condition $\operatorname{Re}\mu>0$ or, equivalently,
\begin{equation}
	\label{Con:1a}
	U > U_c^{(1)}(g)
	= \frac14\left(g^2-\frac{1}{g^2}\right),
\end{equation}
where we have introduced the parameter $g=c/A$.

The second criterion is the condition $\operatorname{Re}E_0>0$. This criterion reflects the fact that, unlike the eigenfunctions of normal operators, the eigenfunction $\Psi_0(\eta)$ of the non-normal operator $\hat{\mathcal{F}}$ can coexist with cloud-function states $\Psi(\eta)$ whose spatial variations, far from the potential well $\Omega(\eta)$, are characterized by length scales much larger than unity. Such states may be associated with eigenvalues close to zero. Transitions between these eigenfunctions are enabled by the nonlinear term in Eq.~\eqref{eq:1DL} (or, equivalently, Eq.~\eqref{eq:1}).

\begin{figure}
	\begin{center}
		\includegraphics[width=0.9\columnwidth]{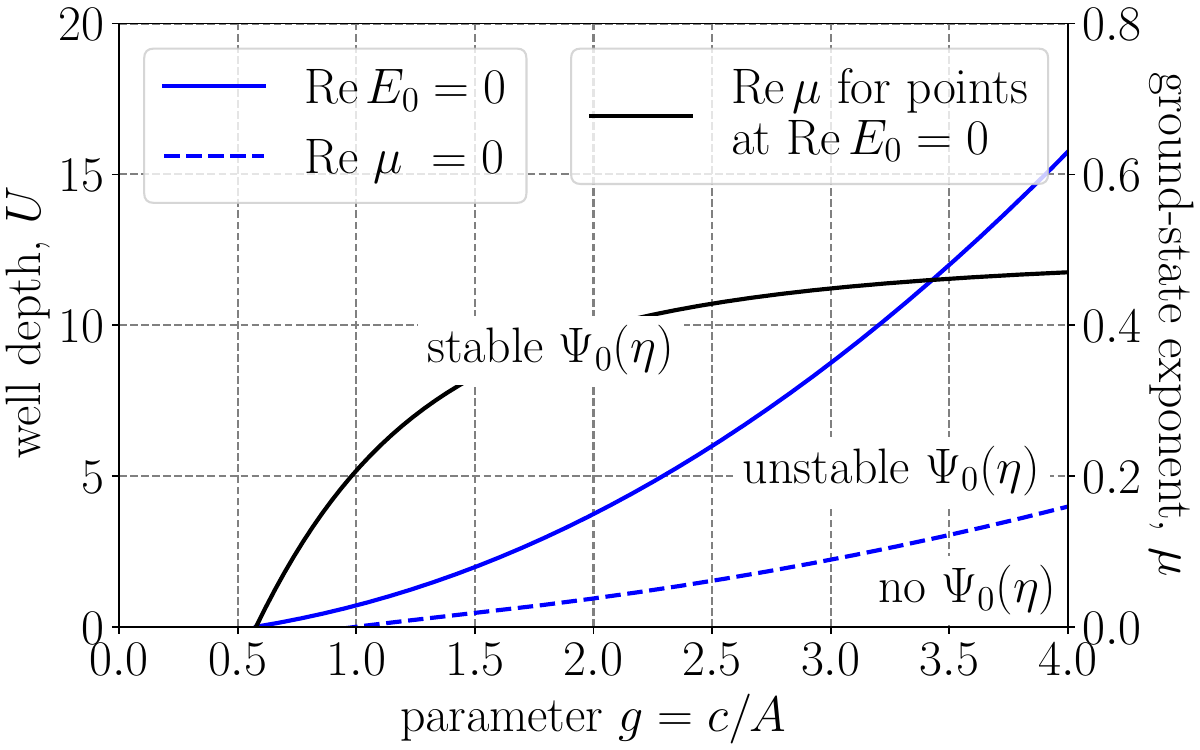}
	\end{center}
	\caption{The phase diagram in the $(U,g)$ parameter plane for the ground-state eigenfunction $\Psi_0(\eta)$. The curves $\operatorname{Re}E_0=0$ and $\operatorname{Re}\mu = 0$ originate at $g = 1/\sqrt{3}$ and $g= 1$, respectively.}
	\label{F:R0}
\end{figure}

Figure~\ref{F:R0} shows the phase diagram in the $(U,g)$ parameter plane obtained numerically. The region bounded from below by the blue solid curve, $\operatorname{Re}E_0=0$, corresponds to the stable ground-state eigenfunction $\Psi_0(\eta)$, namely,
\begin{gather}
	\label{Con:2}
	U > U_c^{(2)}(g),
	\quad\Longleftarrow\quad \operatorname{Re}E_0(U,g)=0.
	\\
	\intertext{Accordingly, the ground-state eigenfunction $\Psi_0(\eta)$ exists but is unstable in the region between the blue solid and dashed curves, i.e., for}
	\label{Con:3}
	U_c^{(1)}(g)<U<U_c^{(2)}(g).
	\\
	\intertext{The ground-state eigenfunction $\Psi_0(\eta)$ is absent below the blue dashed curve, i.e., for}
	\label{Con:4}
	U<U_c^{(1)}(g).
\end{gather}
The ordering of the curves is justified by evaluating $\operatorname{Re}\mu$ (black line) along the locus $\operatorname{Re}E_0=0$.

Accordingly, three distinct regimes of sensory information processing can be identified, depending on the values of the attention degree $A$ and the well depth $U$, reflecting the structure of the SPC hierarchy.
\begin{enumerate}[topsep=0.1\baselineskip,itemsep=0.1\baselineskip,label=\Roman*.]
	\item \textit{Subliminal processing.}\par
	When $U<U_c^{(2)}(c)$, the external stimulus is too weak for the corresponding fragment of the GNW landscape---the potential well~\eqref{eq:Om}---to support the emergence of its high-level representation in the GNW. 
	
	\item \textit{Supraliminal unattended processing.}\par
	When $U>U_c^{(2)}(c)$ but $A<A_c$, where the critical value $A_c$ is specified by the condition
	\begin{equation}
		\label{Con:1b}
		U  = U_c^{(2)}\left(\frac{c}{A_c}\right),
	\end{equation}
 	the stimulus is sufficiently strong to potentially give rise to a representation in the GNW, but insufficient top-down attention prevents its emergence.
	
	\item \textit{Supraliminal attended processing.}\par
	When $U>U_c^{(2)}(c)$ and $A>A_c$, the stimulus is sufficiently strong to generate a representation in the GNW, and top-down attention enables its emergence.
\end{enumerate}

It is worth noting that the emergence of a stimulus representation in the GNW, as the attention degree $A$ increases and exceeds the threshold $A_c$, can be regarded as a first-order phase transition. Indeed, it corresponds to a transition between unstable and stable states of the cloud function $\Psi_0(\eta)$, whose shape is already fixed and remains essentially unchanged in the vicinity of $A_c$. In particular, the emerging high-level representation is initially characterized by a finite extension in the state space $\mathbb{R}_\eta$. By contrast, the emergence of bound states in Hermitian potential wells can be viewed as analogous to a second-order phase transition, since at the binding threshold the localization length diverges and the bound-state energy approaches the continuum edge.

The dynamics of the cloud function $\Psi$ was investigated numerically, and the results are presented in the next section.

\section{Numerical Results and Discussion}

\begin{figure*}[p]
	\begin{center}
		\begin{minipage}{0.75\textwidth}
			\includegraphics[width=\textwidth]{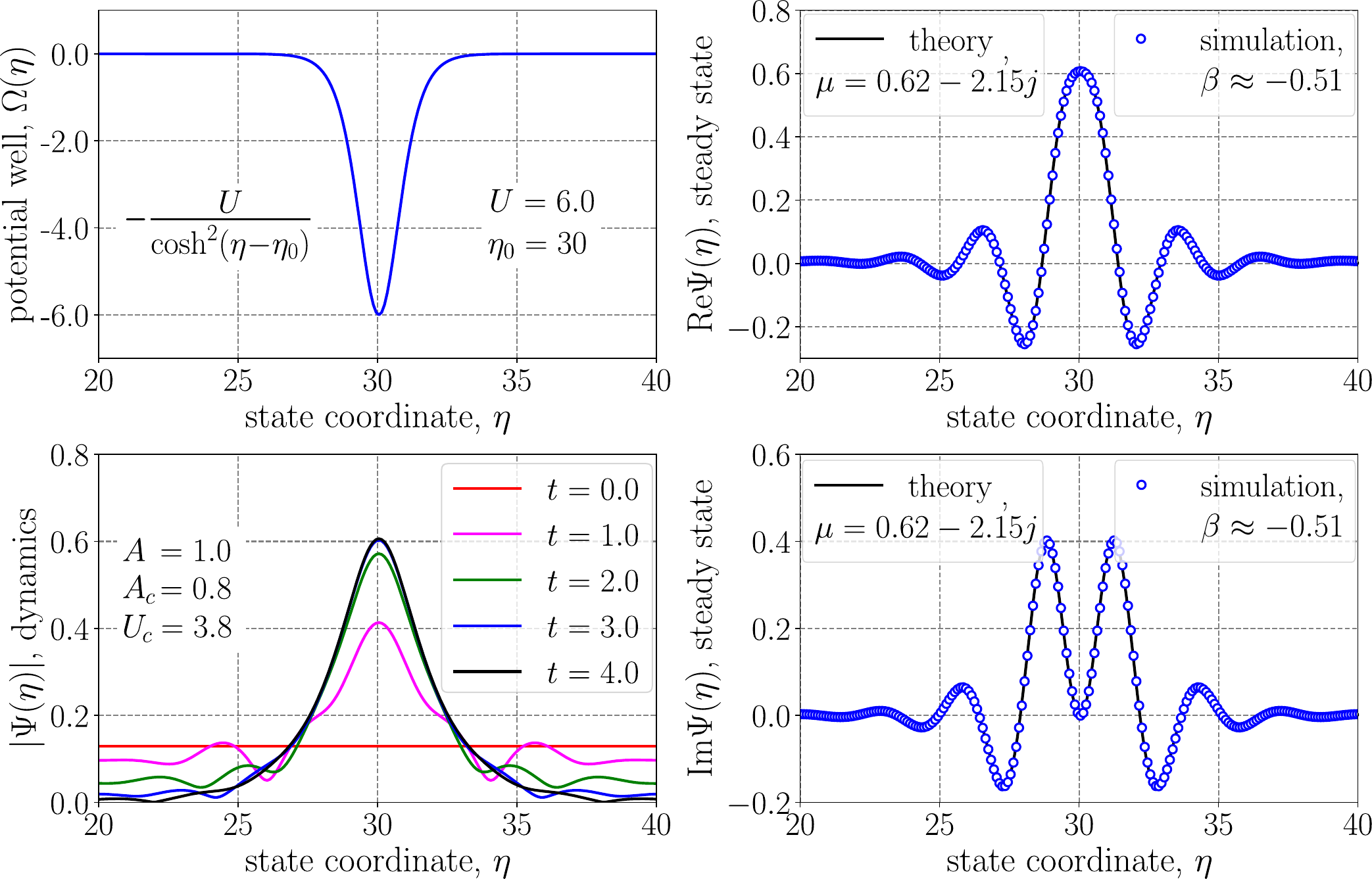}
			\caption{Illustration of the cloud-function dynamics corresponding to the supraliminal attended processing regime.}
			\label{F:R1}
			\vspace*{\baselineskip}
			\includegraphics[width=\textwidth]{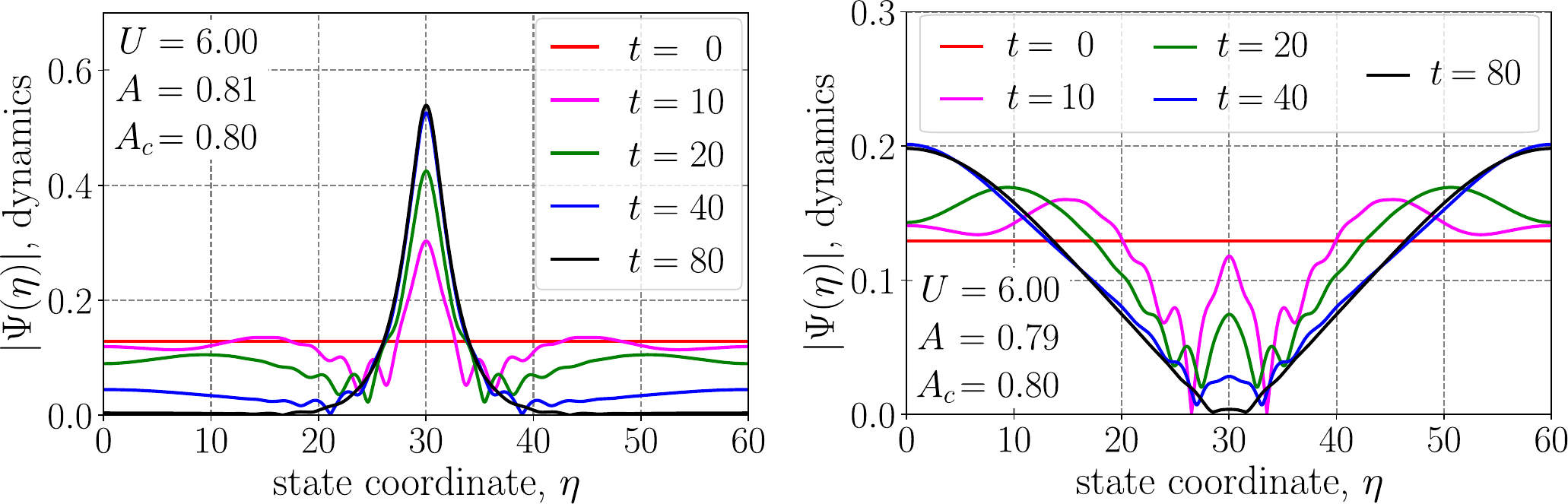}
			\caption{Illustration of the stepwide emergence of stimulus representations in the GNW when the attention degree $A$ exceeds the threshold $A_c$.}
			\label{F:R2}
			\vspace*{\baselineskip}
			\includegraphics[width=\textwidth]{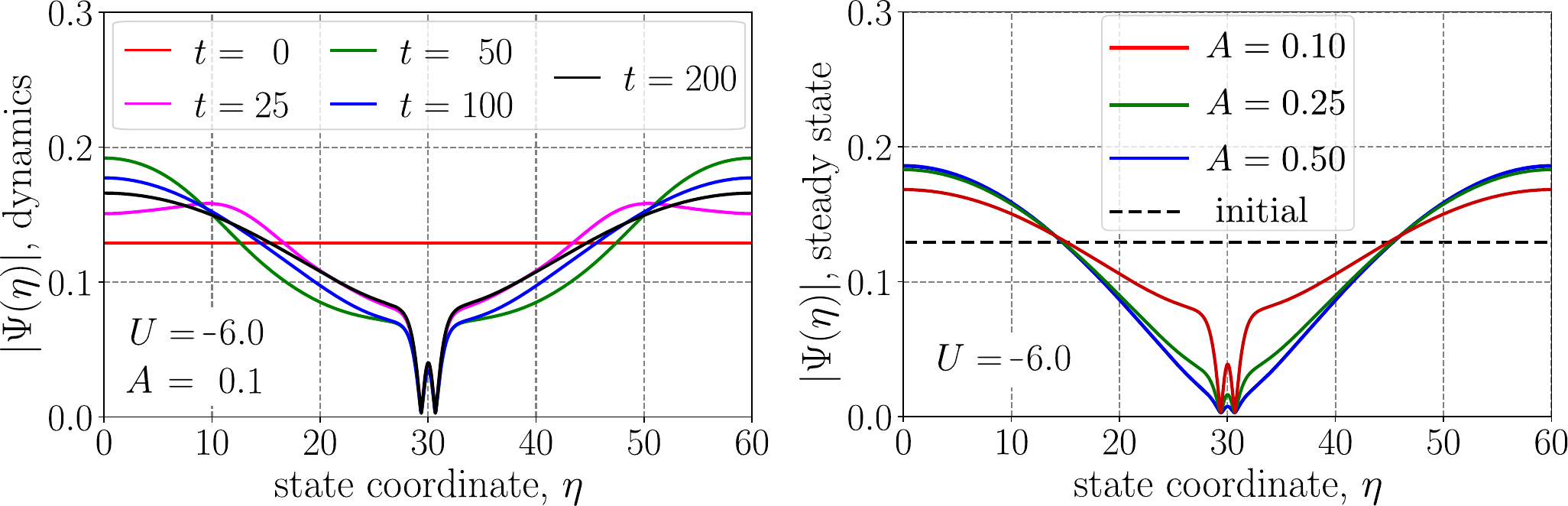}
			\caption{Illustration of the cloud-function dynamics near a maximum of the GNW landscape, modeled by the potential~\eqref{eq:Om} with the sign of its amplitude reversed, $U\to -U$.}
			\label{F:R3}
		\end{minipage}	
	\end{center}
\end{figure*}

The governing equation~\eqref{eq:1DL} was solved numerically using the second-order implicit Crank--Nicolson (CN) scheme combined with the second-order explicit Adams--Bashforth (AB2) method for the nonlinear term. The control over the cloud-function normalization~\eqref{IP:2} was also implemented. The cloud function $\Psi$ was discretized on the interval $[0,L]$ with $L=60$, subject to periodic boundary conditions. The resulting cyclic tridiagonal systems arising from the CN discretization were solved using the periodic Thomas algorithm. The spatial and temporal discretization parameters were set to $dx=0.005$ and $dt=0.001$, respectively. The potential well~\eqref{eq:Om} was centered within the computational domain. The parameter $c$ was set to $c=2$ throughout all simulations.

\subsection{Stability of the ground state and supraliminal attended processing}

Figure~\ref{F:R1} illustrates the evolution of the cloud function $\Psi(\eta,t)$ from the uniform distribution over the computational domain at $t=0$ to the ground-state eigenfunction $\Psi_0(\eta)$, Exp.~\eqref{eq:F2}, which is stable for the well depth $U>U_c$ when the attention degree exceeds its critical value $A_c < 1$. In the simulations considered here, the attention degree is set to its maximum value, $A=1$.

Since the Hamiltonian $\hat{\mathcal{F}}$ is non-normal, its eigenfunctions are generally non-orthogonal. Consequently, an analytical proof that the nonlinear equation~\eqref{eq:1DL} indeed describes the ``winner-takes-all'' competition among the  eigenfunctions requires a dedicated analysis. In the present work, we restrict ourselves to a numerical verification of this property. The right column of Fig.~\ref{F:R1} compares the obtained steady-state cloud function $\Psi_{\mathrm{ss}}(\eta)$ with the analytically constructed eigenfunction $\Psi_0(\eta)$ for $\operatorname{Im} Z_0=0$. As can be seen, the two functions coincide up to a phase factor $e^{i\beta}$, reflecting the arbitrariness of the eigenfunction phase. The fitted value of $\beta$ and the exponent $\mu$ corresponding to the chosen parameter values are also shown in Fig.~\ref{F:R1}.

\subsection{Emergence of high-level representations in the GNW}

In Sec.~\ref{MD}, we proposed that stimulus representations emerge in the GNW through a stepwise transition once the attention degree $A$ exceeds the threshold value $A_c$. This prediction is confirmed numerically, as illustrated in Fig.~\ref{F:R2}. As can be seen, when the attention degree is set to $A = 0.81$, exceeding the threshold $A_c = 0.8$ by $1.25\%$, the resulting bound state of $\Psi$ is nearly identical to that obtained for $A = 1$ (Fig.~\ref{F:R1}), with all other parameters kept unchanged. By contrast, when the attention degree is reduced to $A = 0.79$, lying $1.25\%$ below the threshold, no bound state is formed.

The form of the delocalized state depends only weakly on the subsequent decrease in the attention degree~$A$. The coexistence of the bound and delocalized states in the region~\eqref{Con:4}, which is responsible for this stepwise transition, also allows us to regard the delocalized state as the ground state of the continuous spectrum of the Hamiltonian~$\hat{\mathcal{F}}$.

\subsection{Maxima of the GNW landscape}

In the previous discussion, we focused on the minima of the GNW landscape, assuming them to act as attractors for the dynamics of the cloud function $\Psi$. In this context, the role of the landscape maxima also requires clarification. Indeed, within the proposed model of a complex-valued landscape, the cloud-function dynamics in the vicinity of a maximum is likewise governed by equation~\eqref{eq:1DL} with the operator $\hat{\mathcal{F}}$, Exp.~\eqref{eq:F1}, after reversing the sign of the potential~\eqref{eq:Om}, i.e., $U\to -U$. In this case, the potential component $-A\Omega(\eta)$ becomes a barrier that forces the cloud function $\Psi$ to avoid the region where $\Omega(\eta)$ is localized, whereas the component $ic\Omega(\eta)$ becomes a potential well. According to Exp.~\eqref{eq:F3}, this landscape maximum also supports bound eigenfunctions. These states, however, must be unstable because the Hermitian component of the operator $\hat{\mathcal{F}}$---which is responsible for dissipation in the dynamics of $\Psi$---induces transitions from ``lower'' to ``higher'' eigenstates of this potential well and ultimately drives the cloud function out of it. This behavior is confirmed by the numerical simulations shown in Fig.~\ref{F:R3}. As can be seen, only the delocalized ground state is stable. Therefore, the maxima of the GNW landscape do not contribute directly to the recognition of external stimuli.

\section{Conclusion}

Based on our previously developed neural field formalism \citep{LubashNatPhen2025,Lubashevskiy2026}, we proposed a phenomenological description of the Global Neuronal Workspace (GNW) as a Hilbert space $\mathbb{H}$ over a state space $\mathbb{R}$, whose points represent possible properties of an observed physical system. The space $\mathbb{H}$ is endowed with a complex-valued landscape emerging during the early stage of sensory processing and governing the dynamics of high-level stimulus representations, i.e., the elements of $\mathbb{H}$. Such a representation---the cloud function $\Psi$---inherits the spatial structure of the corresponding mental image, the space--time cloud \citep{LubPlav2021}, while its dynamics is determined by the underlying neural processes. The minima of the real part of the GNW landscape act as attractors of the cloud-function dynamics.

The constructed governing equation resembles a nonlinear Schr\"odinger equation in imaginary time with a non-Hermitian, non-normal Hamiltonian supplemented by a nonlinear term that preserves the cloud-function norm while allowing spatially nonlocal interactions in~$\mathbb{H}$. Owing to the specific structure of the evolution equation, rather than the Hermiticity properties of the Hamiltonian alone, the Hermitian and anti-Hermitian components are assumed to describe complementary aspects of sensory processing in the GNW. The Hermitian component generates a dissipative dynamics that drives the cloud function toward a minimum of the GNW landscape and is interpreted as the recognition process. By contrast, the anti-Hermitian component promotes spreading of the cloud function over the state space~$\mathbb{R}$, which is related to information broadcasting across the GNW.

Analytical and numerical studies demonstrated that the proposed formalism reproduces the tripartite taxonomy of sensory processing, comprising subliminal, preconscious, and conscious regimes. In particular, we showed that the emergence of a bound cloud-function state at a landscape minimum requires two necessary conditions to be satisfied: both the landscape depth and the degree of top-down attention must exceed their respective threshold values.

The previous applications of this neural-field formalism to explaining the power law of working memory and the change-of-mind phenomenon \citep{LubashNatPhen2025,Lubashevskiy2026}, together with the present reproduction of the tripartite taxonomy, suggest that the proposed approach provides a promising framework for describing cognitive phenomena from the first-person perspective. A natural direction for further development is to account for an important feature of sensory processing in the GNW that is not captured by the current formalism, namely the persistence of large-scale reverberatory activity after the external stimulus has disappeared. Within the proposed framework, this effect could be incorporated naturally by generalizing the non-Hermitian Hamiltonian $\hat{\mathcal{H}}$ to include an additional nonlinear term proportional to a power of the cloud function, for example, $|\Psi|^2$. Such an extension would bring the formalism closer to classical field theories based on the complex Ginzburg--Landau equation.

\section*{CRediT authorship contribution statement}

\textbf{V. Lubashevskii}:  Writing --- review \& editing, Conceptualization, Methodology, Software.

\textbf{I. Lubashevsky}: Writing --- original draft, Simulation, Visualization.

\section*{Declaration of competing interest}

The authors declare that they have no known competing financial interests or personal relationships that could have appeared to influence the work reported in this paper.

\bibliographystyle{elsarticle-num} 

\bibliography{%
	../../Bibliography/Time-Physics,%
	../../Bibliography/PhilosophySocialSystems,%
	../../Bibliography/library,%
	../../Bibliography/TrafficPhysics,%
	../../Bibliography/Mathematics,%
	../../Bibliography/Physics,%
	../../Bibliography/Motor-Behavior,%
	../../Bibliography/Psychology,%
	../../Bibliography/PhysiologyofHumans,%
	../../Bibliography/bookPhilosophy,%
	../../Bibliography/Brain,%
	../../Bibliography/photos,%
	../../Bibliography/Categorization,%
	../../Bibliography/MyWorks,%
	../../Bibliography/PhysicsofMind,%
	../../Bibliography/Mind-Phenomenology,%
	../../Bibliography/Learning,%
	../../Bibliography/Memory,%
	../../Bibliography/Consciousness,%
	../../Bibliography/Anticipation,%
	../../Bibliography/Emotions,%
	../../Bibliography/Imagination,%
	../../Bibliography/Decision_Making,%
	../../Bibliography/ActionAgency,%
	../../Bibliography/AttentionAction,%
  	../../Bibliography/HumanIntermittency,%
  	../../Bibliography/ReactionDelay_ActionStimulation,%
  	../../Bibliography/MentalSpaceTime,
  	../../Bibliography/Blur,%
  	../../Bibliography/Cross_Modal,%
	../../Bibliography/Mental-Time-Travels,%
	../../Bibliography/Book_3,%
	../../Bibliography/Events}


\end{document}